\documentclass[12pt]{iopart}

\usepackage{iopams}
\usepackage{graphicx}
\begin{document}

\title[Intermediate valence systems CeRhSi$_2$ and Ce$_2$Rh$_3$Si$_5$]{Magnetic properties and electronic structures of intermediate valence systems CeRhSi$_2$ and Ce$_2$Rh$_3$Si$_5$}

\author{D.~Kaczorowski,$^1$ A.~P.~Pikul,$^1$ U.~Burkhardt,$^2$ M.~Schmidt,$^2$
A.~\'Slebarski,$^3$ A.~Szajek,$^4$ M.~Werwi{\'n}ski,$^4$ and Yu.~Grin$^2$}

\address{$^1$Institute of Low Temperature and Structure
Research, Polish Academy of Sciences, P. O. Box 1410, 50--950 Wroc{\l}aw, Poland \\ $^2$Max Planck
Institute for Chemical Physics
of Solids, N{\"o}thnitzer Str.~40, 01187 Dresden, Germany \\
$^3$Institute of Physics, University of Silesia, 40--007 Katowice, Poland \\ $^4$Institute of
Molecular Physics, Polish Academy of Sciences, M.~Smoluchowskiego 17, 60--179 Pozna{\'n}, Poland}

\ead{D.Kaczorowski@int.pan.wroc.pl}

\begin{abstract}

The crystal structures and the physical (magnetic, electrical transport and thermodynamic)
properties of the ternary compounds CeRhSi$_2$ and Ce$_2$Rh$_3$Si$_5$ (orthorhombic CeNiSi$_2$- and
U$_2$Co$_3$Si$_5$-type structures, respectively) were studied in wide ranges of temperature and
magnetic field strength. The results revealed that both materials are valence fluctuating systems,
in line with previous literature reports. Direct evidence for valence fluctuations was obtained by
means of Ce L$_{\rm III}$-edge x-ray absorption spectroscopy and Ce 3d core-level x-ray
photoelectron spectroscopy. The experimental data were confronted with the results of
\textit{ab-initio} calculations of the electronic band structures in both compounds.

\end{abstract}


\submitto{\JPCM}

\maketitle

\section{Introduction}
The system Ce--Rh--Si is known to be exceptionally rich in ternary phases, which scan a full
spectrum of magnetic behavior related to the hybridization of cerium $4f$ electronic states with
$d$ and $p$ states of neighboring atoms. The most intensively studied has been so far the
tetragonal compound CeRh$_2$Si$_2$ that is considered as an archetypal example of pressure-induced
superconductivity emerging from magnetically ordered state \cite{review}. At ambient pressure, the
$4f$ electrons in CeRh$_2$Si$_2$ are well localized and give rise to an antiferromagnetic ordering
setting in at $T_{\rm N1}$ = 36 K that is followed by a change of the magnetic structure at $T_{\rm
N2}$ = 25 K \cite{122normal}. Under high pressure one observes a gradual suppression of the
N{\'e}el temperature down to absolute zero in the vicinity of the critical pressure $p_{\rm c}
\approx$ 1 GPa \cite{122pres1}. Most interestingly, in the quantum critical region near $p_{\rm c}$
the compound becomes superconducting below $T_{\rm c}$ = 0.5 K \cite{122pres1,122pres2}.
Simultaneously, the Fermi surface topology varies abruptly and the cyclotron mass measured in de
Haas -- van Alphen experiment increases rapidly, hence manifesting a change in character of the
$4f$ electrons, which become itinerant in the critical region \cite{122dhva}. Another
pressure-induced superconductor among the Ce--Rh--Si ternaries is CeRhSi$_3$ \cite{review}. At
ambient pressure, this tetragonal compound orders antiferromagnetically at $T_{\rm N}$ = 1.6 K and
behaves as a Kondo lattice with the characteristic temperature $T_{\rm K}$ of about 100 K
\cite{113normal}. In contrast to the case of CeRh$_2$Si$_2$, in an applied pressure $T_{\rm N}$
does not change monotonically: it first increases up to 1.9 K near 0.8 GPa and then decreases down
to 1.1 K near 2.6 GPa without further change at higher pressures \cite{113pres2}. Most importantly,
superconductivity appears in this compound already under small pressures and coexists with the
antiferromagnetic ordering in the entire pressure range studied, with the critical temperature
gradually rising up to 1.1 K with increasing pressure \cite{113pres2}. Above 2.6 GPa, $T_{\rm c}$
merges with $T_{\rm N}$ \cite{113pres1, 113pres2}. Moreover, in the vicinity of this pressure,
pronounced non-Fermi liquid features are observed in all the bulk properties. The occurrence of
superconductivity in CeRhSi$_3$ is highly peculiar, as the crystal structure of this compound lacks
an inversion symmetry \cite{113pres2}. Apparently, CeRhSi$_3$ is another member of a novel family
of unconventional non-centrosymmetric superconductors with the most prominent representative being
CePt$_3$Si \cite{bauer}. Likely, mixed spin-triplet pairing symmetry in CeRhSi$_3$ is responsible
for extremely large upper critical magnetic field $B_{\rm c2}(0)$ = 30 T measured in a magnetic
field applied along the crystallographic $c$-axis, its very unusual temperature dependence, as well
as strong anisotropy of $B_{\rm c2}(0)$ (cf. $B_{\rm c2}(0)$ = 7 T for $B \perp c$-axis)
\cite{113sup}.

Motivated by the intriguing physics in CeRh$_2$Si$_2$ and CeRhSi$_3$, we started systematic
investigations of the physical behavior in other ternaries from the Ce--Rh--Si phase diagram.
Recently, we reported on very complex magnetic properties of the orthorhombic compound
CeRh$_3$Si$_2$, marked by two subsequent antiferromagnetic phase transitions at $T_{\rm N1}$ = 4.7
K and $T_{\rm N2}$ = 4.5 K, multiple metamagnetic transitions in the ordered state, and huge
magnetocrystalline anisotropy being quite uncommon for Ce-based intermetallics \cite{132}. In turn,
for hexagonal Ce$_2$RhSi$_3$ we provided evidence of Kondo lattice behavior with the characteristic
temperature $T_{\rm K} \approx$ 9 K, which coexists below $T_{\rm N}$ = 4.5 K with long-range
antiferromagnetic ordering \cite{213}. Remarkably, by means of bulk and spectroscopic measurements,
the cerium $4f$ electrons were found to be well localized in both compounds.

An entirely opposite character of the $4f$ states may be expected for the silicides CeRhSi$_2$ and
Ce$_2$Rh$_3$Si$_5$, which have been reported in the literature as valence fluctuating systems
\cite{112adr,112chev,235god,235ram}. The former compound crystallizes with an orthorhombic crystal
structure of the CeNiSi$_2$-type. Its unit cell volume does not follow the lanthanide contraction
established for the $RE$RhSi$_2$ ($RE$ = La, Pr, Nd) series. The magnetic susceptibility of
CeRhSi$_2$ shows a non-Curie-Weiss temperature variation with a broad maximum near 80 K. The
electrical resistivity is proportional to the squared temperature below about 50 K, while above 100
K it decreases slightly with increasing temperature. These features of the bulk magnetic and
electrical behaviors reflect instability of the Ce ions' valence. The Coqblin-Schrieffer approach
applied to the magnetic data of CeRhSi$_2$ yielded the characteristic temperature $T_0$ = 309 K
($T_0$ is related to the Kondo temperature $T_{\rm K}$ via the Wilson number, $T_{\rm K} = WT_0$).
In turn, analysis of the resistivity data in terms of the Freimuth model gave estimates for the
spin-fluctuation temperature $T_{\rm sf}$ = 147 K and the position of the $4f$ band with respect to
the Fermi energy $T_{\rm f}$ = 30 K. The valence fluctuating nature of CeRhSi$_2$ has also been
concluded in an independent study reported in Ref. \cite{112chev}. The other compound of our
interest, i.e. Ce$_2$Rh$_3$Si$_5$, was studied in less details. It was reported to form with the
orthorhombic U$_2$Co$_3$Si$_5$-type structure as a member of the $RE$$_2$Rh$_3$Si$_5$ ($RE$ =
La--Er) family. As the lattice parameters of its unit cell follow the lanthanide contraction along
the series it was concluded in Ref. \cite{235god} that the Ce ions in this compound are trivalent.
On the other hand, the same authors reported nearly temperature independent magnetic susceptibility
of Ce$_2$Rh$_3$Si$_5$, and in order to reconcile their contradicting findings they speculated that
the ground state in this compound is nonmagnetic because of the presence of 'virtual' spin
fluctuations, which however do not destabilize the charge state of the Ce ions. The electrical
transport properties of Ce$_2$Rh$_3$Si$_5$ were reported in Ref. \cite{235ram}. In the temperature
range 2-30 K, the resistivity was shown to vary as $T^3$, and interpreted as being governed by
interband $s-d$ scattering. Nonmagnetic character of Ce$_2$Rh$_3$Si$_5$ was also inferred from the
heat capacity data, which show featureless temperature behavior up to 30 K and strongly reduced
entropy $S(30 \rm K) = 0.24R$.

In this work we reinvestigated the magnetic and electrical transport properties of CeRhSi$_2$ and
Ce$_2$Rh$_3$Si$_5$. Moreover, we measured for the first time the specific heat of both materials in
a wide temperature range 0.35-300 K, and performed x-ray absorption spectroscopy (XAS) and x-ray
photoelectron spectroscopy (XPS) studies. The experimental data are discussed in the context of the
calculated electronic structures. Altogether, the obtained results confirm the valence fluctuating
nature of both compounds.

\section{Experimental and computational details}
Polycrystalline samples of CeRhSi$_2$ and Ce$_2$Rh$_3$Si$_5$ were prepared by arc melting
stoichiometric amounts of the elemental components (Ce - 3N, Ames Laboratory, Rh - 3N, Chempur and
Si - 6N, Chempur) in a copper-hearth furnace installed inside a glove-box filled with ultra-pure
argon gas with continuously controlled partial pressures of O$_2$ and H$_2$O to be lower than 1
ppm. The buttons were flipped over and remelted several times to ensure good homogeneity. The
weight losses after the final melting were negligible (less than 0.2\%).

The quality of the obtained alloys was checked by x-ray powder diffraction on an X'pert Pro
PanAnalytical diffractometer with CuK$_\alpha$ radiation and by energy dispersive x-ray (EDX)
analysis using a Phillips 515 scanning electron microscope equipped with an EDAX PV 9800
spectrometer. Both techniques proved single-phase character of the Ce$_2$Rh$_3$Si$_5$ sample, with
the expected stoichiometry and crystal structure. In the case of CeRhSi$_2$, however, some small
admixture of CeRh$_2$Si$_2$ was evidenced in the x-ray pattern and the EDX spectrum. The structural
refinements were done employing the program FULLPROF \cite{fullprof}.

Magnetic susceptibility measurements were performed in the temperature range 1.72-800 K in magnetic
fields of 0.5 T using a Quantum Design superconducting quantum interference device (SQUID)
magnetometer. The heat capacity and the electrical resistivity were measured within the temperature
interval 2-300 K using a Quantum Design PPMS platform.

The x-ray photoelectron spectroscopy experiments were carried out at room temperature using a
Physical Electronics PHI 5700/660 spectrometer with monochromatized AlK$\alpha$ radiation (1486.6
eV). The spectra were collected on parallelepiped-shaped specimens broken in-situ in high vacuum of
the order of $10^{10}$ Torr.

The measurements of the x-ray absorption spectroscopy at the Ce-L$_{\rm {III}}$ threshold were
performed at several different temperatures at the EXAFS-1 beamline C of the Hamburger
Synchrotronstrahlungslabor (HASYLAB/DESY) using a fixed-exit double-crystal Si (111) monochromator.
In these studies CeO$_2$ and CeF$_3$ were used as the internal standards. The energy scans were
made step-by-step with an energy resolution of approx. 1 eV. The absorption spectra were calculated
by $ln(C_1/C_2)$ from the x-ray intensities $C_1$, $C_2$, detected by ionization chambers in front
and behind the irradiated flat sample. A peak to background ratio $P/B$ = 0.6 at the Ce-L$_{\rm
{III}}$ white line was realized with powdered samples of about 16 mg, which were ground with small
amounts of B$_4$C and fixed with paraffin wax on a 1 cm$^2$ window of the flat copper sample
holder. Temperatures from the range 5-293 K were obtained by a He-gas flow cryostat showing thermal
stabilities of $T = +/- 0.5$ K during typical measuring dwell times of 20 minutes.

The electronic structure calculations were performed within density functional theory \cite{as01}
using the full-potential local-orbital minimum-basis band structure scheme (FPLO) \cite{as02} and
the full potential linearized augmented plane wave (LAPW) method implemented in the latest version
(WIEN2k) of the original WIEN code \cite{as03}. The FPLO calculations were performed in the
fully-relativistic mode and the LSDA (the local spin density approximation) exchange-correlation
potential was assumed in the form proposed by Perdew and Wang \cite{as04}. In the Wien2k code
calculations the scalar relativistic approach was implemented with the spin-orbit interactions
taken into account using the second variational method \cite{as05}. Two different
exchange-correlation potentials in the generalized gradient approximations (GGA) were tested, in
the forms proposed by Perdew \emph{et al.} \cite{as06} as well as Wu and Cohen \cite{as07}.
Furthermore, to improve the description of the strongly correlated 4$f$ electrons, the on-site
Coulomb energy U correction was introduced within the LSDA+U approach \cite{as08}. The value of
U$_{eff}$ was chosen equal to 6 eV \cite{as09}. The number of k-points was 8000 in the Brillouin
zone (BZ), which corresponds to at least 1100 points in the irreducible wedge of the BZ for all
systems and methods of calculations. For BZ integration a tetrahedron method was used \cite{as10}.
The self-consistent criterion was equal to at least 10$^{-6}$ Ry for the total energy. The
calculations were performed for lattice constants and atomic positions in the unit cells as given
in the IIIA section.

The theoretical photoemission spectra (XPS) were obtained from the calculated densities of
electronic states (DOS) convoluted by Gaussian with a half-width ($\delta)$ equal to 0.3 eV and
scaled using the proper photoelectronic cross sections for partial states \cite{as11}.

\section{Results and discussion}
\subsection{Crystal structures}
The compound CeRhSi$_2$ was reported in the literature to crystallize with an orthorhombic
structure of the CeNiSi$_2$-type \cite{112adr,112chev}. In turn, Ce$_2$Rh$_3$Si$_5$ was considered
as having an orthorhombic structure isotypic to U$_2$Co$_3$Si$_5$ \cite{235god,235ram}. However, to
the best of our knowledge, for none of these two phases were any details on the atomic positions in
the unit cells reported. Therefore, the x-ray powder diffraction data obtained in the present study
as part of characterization of the samples quality were used to refine the crystal structures of
both compounds.

For CeRhSi$_2$, the refined lattice parameters are $a$ = 4.2615(3) \AA, $b$ = 16.7469(9) \AA and
$c$ = 4.1751(3) \AA, in good agreement with those given in Ref. \cite{112chev}. The crystal
structure belonging the space group $Cmcm$ was refined down to the residuals $R_{\rm p}$ = 3.4\%
and $R_{\rm {wp}}$ = 5.3\%. The obtained atomic coordinates are given in Table 1, together the
values of equivalent isotropic thermal displacement parameters for all the atoms. The compound was
thus confirmed to be isostructural with CeNiSi$_2$. Detailed discussion of this crystal structure
can be found in the literature, e.g. in the original paper \cite{cenisi2}. For the purpose of this
work it is enough to note that the unit cell contains one position of Ce atoms, which are
coordinated by four Rh atoms at a distance of 3.226 \AA, four Si1 atoms at a distance of 3.163 \AA,
and two Si2 atoms at a distance of 3.167 \AA.

As for Ce$_2$Rh$_3$Si$_5$, the unit cell of the U$_2$Co$_3$Si$_5$-type  (space group $Ibam$) has
also been corroborated. The crystal structure was refined down to the residuals $R_{\rm p}$ = 2.3\%
and $R_{\rm {wp}}$ = 3.1\%. The obtained lattice parameters are $a$ = 9.8949(3) \AA, $b$ =
11.7576(3) \AA and $c$ = 5.8114(1) \AA. These values differ only slightly from those reported in
Refs. \cite{235god,235ram}. The atomic coordinates and the equivalent isotropic thermal
displacement parameters are collected in Table 2. In the unit cell of Ce$_2$Rh$_3$Si$_5$ there is
one position of Ce atom that is surrounded by one Rh2 atom at a distance of 3.138 \AA, two Rh2
atoms at a distance of 3.161 \AA, one Rh2 atom at a distance of 3.375 \AA, two Rh1 atoms at a
distance of 3.381 \AA and one Rh2 atom at a distance of 3.399 \AA. The nearest neighbors silicon
atoms are located at the distances: 3.022 \AA (one Si3 atom), 3.111 \AA (two Si3 atoms), 3.129 \AA
(one Si1 atom), 3.148 \AA (one Si3 atom), 3.199 \AA (two Si2 atoms) and 3.233 \AA (one Si2 atom).
Further details on the U$_2$Co$_3$Si$_5$-type crystal structure can be found in the literature (see
e.g. Ref. \cite{235ram}).

\subsection{Magnetic properties}
The magnetic data of CeRhSi$_2$ and Ce$_2$Rh$_3$Si$_5$ are summarized in Fig. 1. The compounds
exhibit small (of the order of $10^{-3}$ emu/mol per Ce atom) and weakly temperature dependent
magnetic susceptibility that clearly manifests nonmagnetic character of both materials. With
decreasing temperature from 800 K the susceptibility of CeRhSi$_2$ slightly increases, goes through
a broad maximum centered around 75 K and then rapidly rises below about 20 K. Similar behavior is
observed for Ce$_2$Rh$_3$Si$_5$, with a maximum located around 270 K and some tendency for
saturation before the low-temperature upturn occurs. The overall shape of these $\chi(T)$ curves is
typical for Ce-based intermetallics with valence fluctuations. The position of the maximum in
$\chi(T)$, $T(\chi^{\rm {max}})$, gives an estimate for the characteristic temperature $T_{\rm
{sf}}$, related to spin fluctuations in such compounds. From the relation \cite{lrp}

\begin{equation}
T_{\rm {sf}}= \frac{3}{2}T(\chi^{\rm {max}})
\end{equation}

one finds $T_{\rm {sf}} \simeq$ 112 K for CeRhSi$_2$ and $T_{\rm {sf}} \simeq$ 405 K for
Ce$_2$Rh$_3$Si$_5$. Comparison of these values suggests that in these two compounds interactions of
the cerium $4f$ electrons with the conduction band is stronger in the latter one, in line with
smaller and less temperature dependent magnetic susceptibility observed for Ce$_2$Rh$_3$Si$_5$.

Low-temperature upturns in $\chi(T)$ curves are commonly observed in Ce-based intermediate valence
materials and their origin is usually attributed to the presence of stable Ce$^{3+}$ ions located
at grain boundaries or/and some contamination by paramagnetic impurities. To account for this
spurious effect it is assumed that the intrinsic susceptibility of CeRhSi$_2$ and
Ce$_2$Rh$_3$Si$_5$ at low temperatures is given by the formula \cite{bl}

\begin{equation}
\chi(0) = \frac{C}{2T_{\rm {sf}}}
\end{equation}

in which $C$ = 0.807 emu/(mol K) stands for the Curie constant of free Ce$^{3+}$ ions ($C =
\frac{N\mu _{\rm {eff}} ^2}{3k_{\rm B}}$ where $N$ is the Avogadro number, $k_{\rm B}$ is the
Boltzmann constant, while $\mu _{\rm {eff}}$ = 2.54 $\mu_B$ is the effective magnetic moment of the
cerium $4f^1$ state). The above-derived values of $T_{\rm {sf}}$ imply $\chi(0)$ equal to $3.6
\cdot 10^{-3}$ emu/mol for CeRhSi$_2$ and $1.0 \cdot 10^{-3}$ emu/mol per Ce atom for
Ce$_2$Rh$_3$Si$_5$. Then, the measured magnetic susceptibility can be modeled by the function

\begin{equation}
\chi(T \rightarrow 0) = \chi(0) + \frac{nC}{T}
\end{equation}

where the Curie term represents the contribution due to spurious Ce$^{3+}$ ions in an amount of $n$
atoms per mole of the given compound. Fitting this equation to the experimental data taken below 20
K (note the dashed lines in Fig. 1) yields $n$ = $6.3 \cdot 10^{-3}$ and $3.5 \cdot 10^{-3}$ for
CeRhSi$_2$ and Ce$_2$Rh$_3$Si$_5$, respectively. The magnetic susceptibility data corrected for the
spurious Ce$^{3+}$ ions is shown in Fig. 1 by the full symbols.

At higher temperatures the magnetic susceptibility of both compounds can be analyzed in terms of
the interconfiguration fluctuations (ICF) model developed for intermediate valence systems by Sales
and Wohlleben \cite{sw}. Within this approach the magnetic susceptibility of a Ce-based compound
with nonmagnetic $4f^0$ ground state is expressed as

\begin{equation}
\chi(T) = \frac{C[1 - \nu (T)]}{T + T_0} + \chi_0
\end{equation}

where $T_0$ is a characteristic temperature associated with valence fluctuations between the $4f^0$
and $4f^1$ configurations of Ce ions, whereas $\nu (T)$ stands for the temperature dependent mean
occupation of the ground state that is given by the formula

\begin{equation}
\nu (T) = \frac{1}{1+6\exp(\frac{-E_{\rm {ex}}}{T + T_0})}
\end{equation}

in which $E_{\rm {ex}}$ denotes the energy difference between the ground and excited states. The
constant term $\chi_0$ in Eq. 4 accounts for paramagnetic contribution due to conduction electrons
and diamagnetic contribution due to core electrons. Fitting the ICF model to the experimental data
of CeRhSi$_2$ and Ce$_2$Rh$_3$Si$_5$ resulted in reasonable descriptions of the magnetic behavior
above about 50 K and 150 K, respectively (note the solid lines in Fig. 1). The parameters obtained
in the least-squares fits are $E_{\rm {ex}}$ = 220 K, $T_0$ = 44 K and $\chi_0 = 2.3 \cdot 10^{-4}$
emu/mol for the former compound, and $E_{\rm {ex}}$ = 845 K, $T_0$ = 129 K and $\chi_0 = 4.9 \cdot
10^{-4}$ emu/mol for the latter one. Assuming that the highest possible valence for the cerium
$4f^0$ state is +3.30 (for a discussion see Ref. \cite{val}), one may estimate from Eq. 5 the
change of the effective valence of Ce ions with varying temperature. In the range in which the ICF
model provides proper approximation of the susceptibility of the compounds studied one finds for
CeRhSi$_2$ the valence change from +3.19 at 50 K to +3.07 at 300 K and to +3.05 at 800 K, while for
Ce$_2$Rh$_3$Si$_5$ the change is from +3.23 at 150 K to +3.16 at 300 K and to +3.09 at 800 K.

\subsection{Heat capacity}
Fig. 2 displays the temperature dependences of the specific heat of CeRhSi$_2$ and
Ce$_2$Rh$_3$Si$_5$. The solid lines through the experimental points are the least-squares fits of
the formula:

\begin{equation}\label{Ctot}
C (T) = C_{\rm el}(T) + C_{\rm ph}(T),
\end{equation}

where the first term is the electron specific heat $C_{\rm el}(T) = \gamma T$, while the second one
represents the phonon contribution in the form \cite{gopal,kittel}:

\begin{equation}\label{Cph}
C_{\rm ph}(T) = \frac{1}{1-\alpha T} \left[ \underbrace{9 R \left( \frac{T}{\Theta_{\rm
D}}\right)^3\int_0^{\Theta_{\rm D}/T} \frac{x^4 e^x}{\left( e^x - 1 \right)^2}dx}_{C_{\rm
ph,D}} + \underbrace{\sum_{i} n_{{\rm E}i} R \left( \frac{\Theta_{\rm Ei}}{T} \right)^2
\frac{e^{\Theta_{\rm Ei}/T}}{\left( e^{\Theta_{\rm Ei}/T}-1\right)^2}}_{C_{\rm ph,E}}
\right].
\end{equation}

Here, $\alpha$ stands for the anharmonic coefficient, $R$ is the gas constant, $C_{\rm ph, D}$
describes the Debye contribution of three acoustic modes (characterized by the Debye temperature
$\Theta_{\rm D}$), and $C_{\rm ph, E}$ represents the Einstein specific heat calculated for $i$
groups with $n_i$ optical branches (characterized by the Einstein temperatures $\Theta_{{\rm
E}i}$). In order to avoid overparametrization of the experimental curves $C(T)$ (Fig.~2), we
reduced the number of the groups of optical branches to $i=2$, which seems to be minimum to
describe satisfactorily the specific heat data. In that case we found that the best results one
obtains assuming multiplicity $n_1=2$ and $n_2=7$ in CeRhSi$_2$, and $n_1=12$ and $n_2=15$ in
Ce$_2$Rh$_3$Si$_5$. The so-obtained explicit form of Eq.~(\ref{Ctot}) applied to the experimental
data above about 15 K yields for CeRhSi$_2$ the fitting parameters: $\Theta_{\rm D}$ = 163 K,
$\Theta_{\rm {E_1}}$ = 139 K, $\Theta_{\rm {E_2}}$ = 324 K, $\gamma$ = 22.4 mJ/(mol K$^2$) and
$\alpha \approx 1.0 \cdot 10^{-6}$ 1/K. In the case of Ce$_2$Rh$_3$Si$_5$ a reasonable fit to the
experimental data was obtained with $\Theta_{\rm D}$ = 147 K, $\Theta_{\rm {E_1}}$ = 198 K,
$\Theta_{\rm {E_2}}$ = 427 K, $\gamma$ = 45.3 mJ/(mol K$^2$) and $\alpha = 2.7 \cdot 10^{-4}$ 1/K.
It is worthwhile emphasizing that the presented model does not reflect complexity of the real
lattice vibrations in the two compounds, and at most gives only basic characteristics of the phonon
spectra.

As can be inferred from the insets to Fig. 2, below about 12~K the experimental specific heat data
can be described by a simple formula:

\begin{equation}
C(T)/T = \gamma + r \frac{1944}{\Theta^3_{\rm D, LT}}T^2,
\end{equation}

where the second term is the conventional $T^3$-Debye law with the low-T Debye temperature
$\Theta_{\rm {D,LT}}$ and $r$ stands for a number of atoms in a formula unit \cite{gopal,kittel}.
The insets to Fig. 2 present the least-squares fits of the latter formula to the measured specific
heat data with the parameters $\gamma$ = 86 mJ/(mol K$^2$) and $\Theta_{\rm {D,LT}}$ = 315 K for
CeRhSi$_2$ and $\gamma$ = 47 mJ/(mol K$^2$) and $\Theta_{\rm D, LT}$ = 375 K for
Ce$_2$Rh$_3$Si$_5$. The obtained values of $\Theta_{\rm {D,LT}}$ are obviously different from those
obtained using previous full-range fittings since in the latter approximation only the acoustic
branches are considered, and thus $\Theta_{\rm {D,LT}}$ reflects an average characteristic
temperature of the phonon spectrum. The obtained $\gamma$ coefficients are in both compounds of the
order typical for systems with valence fluctuations \cite{ww}.

At low temperatures approaching the experimental limit a little upturn is observed in $C(T)/T$ of
CeRhSi$_2$. This anomaly has an unknown origin and can temporarily be ascribed to the impurity
phase CeRh$_2$Si$_2$, detected on the x-ray diffraction pattern of the sample measured. However, an
intrinsic nature of the upturn can neither be excluded.

\subsection{Electrical resistivity}
The temperature variations of the electrical resistivity of CeRhSi$_2$ and Ce$_2$Rh$_3$Si$_5$ are
shown in Fig. 3. For both compounds the absolute magnitude of the resistivity is very high because
of many cracks present in the specimens measured (the samples were very brittle). Therefore,
quantitative discussion of the electrical behavior of these silicides is not possible.
Nevertheless, it is clear that both materials exhibit metallic-like conductivity with some features
characteristic of valence fluctuations systems \cite{ww}. Below 50 K for CeRhSi$_2$ and 90 K for
Ce$_2$Rh$_3$Si$_5$ K the resistivity is proportional to the squared temperature, as predicted for
such systems within the paramagnon model \cite{lrp} (some deviations from $\rho \sim T^2$ are
observed below 10 K and 20 K, respectively). At higher temperatures the $\rho (T)$ curve of
CeRhSi$_2$ forms a broad shallow maximum above 100 K, while that of Ce$_2$Rh$_3$Si$_5$ shows a
tendency to saturate near room temperature. The observed behavior is thus fully consistent with the
values of the spin fluctuation temperature $T_{\rm {sf}} \simeq$ 112 K and $T_{\rm {sf}} \simeq$
405 K estimated from the magnetic susceptibility data for the former and the latter compound,
respectively.

\subsection{XAS spectra}
The x-ray absorption spectroscopy data obtained for CeRhSi$_2$ and Ce$_2$Rh$_3$Si$_5$ are gathered
in Fig.~4. For each compound a linear baseline determined at 100 eV below the Ce-L$_{\rm III}$
absorption edge was subtracted, and the spectrum was normalized to the intensity values at $E$ =
5755 eV. The absorption spectra of both phases show one dominant white line at 5726 eV of the
Ce-L$_{\rm III}$ line and several maxima with energies higher than 5750 eV in the extended x-ray
absorption fine structure (EXAFS) region. The EXAFS signal is influenced by specific features of
the crystal structure of the investigated compound. Comparison of the measured spectrum with that
of the Ce$^{3+}$ reference system CePO$_4$ allows us to attribute the observed white line to the
same dominant $4f^1$ electronic configuration of the cerium atoms. Both phases, i.e. CeRhSi$_2$ and
Ce$_2$Rh$_3$Si$_5$, thus show a mean valence of cerium ions relatively close to +3. However, the
presence of an additional contribution at nearly $\Delta E \approx$ 9 eV above the white line
maximum clearly shows mixed valent behavior of the compounds investigated. Decreasing the
temperature from room temperature to $T$ = 5 K leads to a small decrease in the intensity of the
white line and simultaneous increase of the other peak. The observed shift of the spectral weight
of the Ce-L$_{\rm III}$ white line to higher energies implies a reduced influence of the $4f$
electron on the $2p \rightarrow nd$ transition. This behavior manifests a gradual gain in the
contribution due to the $4f^0$ configuration, and thus reflects an increase of the mean valence of
the Ce atoms with decreasing temperature.

In quantitative evaluation of the spectra measured for CeRhSi$_2$ and Ce$_2$Rh$_3$Si$_5$, two
Gaussian functions were considered, which represent the $4f^1$ and $4f^0$ contributions,
respectively (see an example presented in panel (c) of Fig. 4). Additionally, an arctan-function
was added in order to account for the contribution due to transitions of the $2p$ electrons to the
conduction band. Analyzing the relative increase of the $4f^0$ peak with decreasing temperature one
obtains for CeRhSi$_2$ the mean valence of Ce ions changing from +3.15 at room temperature to +3.17
at 5 K (see Fig. 4d). Similar calculation performed for Ce$_2$Rh$_3$Si$_5$ yields the valence
increase from +3.18 at 280 K to +3.22 at 5 K (cf. Fig. 4b).

\subsection{XPS spectra}
The XPS spectra of the 3$d$ core levels usually provide some detailed information about the 4$f$
shell configurations and the $f$-conduction-electron hybridization. The calculations by Gunnarsson
and Sch\"onhammer \cite{GS} indicate that a large hybridization potential, $V_{fs}$, is necessary
to explain the observed spectra for several Ce-intermetallic compounds, which often show different
final states depending on the occupation of the $f$ shell: $f^0$, $f^1$, and $f^2$ (see Ref.
\cite{Fuggle}). Due to the spin-orbit (SO) interaction there are two sets of Ce 3$d$ photoemission
lines in the spectrum attributed to the $3d_{3/2}$ and $3d_{5/2}$ components of the final states,
with an intensity ratio $I(3d_{5/2})/ I(3d_{3/2})=3/2$. The main photoemission lines originating
from Ce$^{3+}$ are labeled as $3d4f^1$.

The  $3d4f^2$ final state components appears when the core hole becomes screened by an additional
$4f$ electron, which is possible due to the hybridization of the Ce $4f$ shell with the conduction
electrons. Consequently, the $3d^9f^2$ components in the Ce $3d$ XPS spectra are attributed within
the Gunnarsson-Sch\"onhammer model to the $f$-conduction electron hybridization energy $\Delta_{\rm
{fs}}$. $\Delta_{\rm {fs}}=\pi V^{2}\rho_{max}$ describes the hybridization part of the Anderson
impurity Hamiltonian \cite{And61}, where $\rho_{max}$ is the maximum in the DOS and V is the
hybridization matrix element. It is possible to estimate $\Delta_{\rm {fs}}$ from the ratio
$r=I(f^{2})/(I(f^{1})+I(f^{2}))$, calculated as a function of $\Delta_{\rm {fs}}$ in Ref.
\cite{Fuggle}, when the peaks of the Ce 3$d$ XPS spectra that overlap are separated.

Fig. 5 shows the Ce 3$d$ XPS spectra of CeRhSi$_2$ ($a$) and Ce$_2$Rh$_3$Si$_5$ ($b$). The
separation of the overlapping peaks in the spectra was done on the basis of  Doniach-\v Sunji\'c
theory \cite{Don70}. A background, calculated using the Tougaard algorithm \cite{Tougaard}, was
subtracted from the XPS data.  The estimated value of the SO splitting equals 18.6 eV. In Fig. 5
each SO set of the Ce $3d$ photoemission lines consists of contributions marked as $f^0$, $f^1$,
and $f^2$. The intensity ratio $r \approx 0.09$ for CeRhSi$_2$ and $\sim 0.17$ for
Ce$_2$Rh$_3$Si$_5$. This intensity ratio gives for CeRhSi$_2$ and Ce$_2$Rh$_3$Si$_5$ a crude
estimate of a hybridization width $\Delta_{\rm {fs}}$ $\sim 40$ meV and  $\sim 83$ meV,
respectively. The estimated energy $\Delta_{\rm {fs}}$ for Ce$_2$Rh$_3$Si$_5$ is twice as large as
the value for CeRhSi$_2$, which suggests a lower occupation number $n_f$  of the $4f$ shell in
Ce$_2$Rh$_3$Si$_5$.

In order to determine the ground state $f$ occupation from the $3d$ XPS spectra we use Fig. 4 and 6
of Ref. \cite{Fuggle}, where the dependence of the ratio $ I(f^{0})/[(I(f^{0})+I(f^{1})+I(f^{2})]$
on the $f$ occupation  is shown for different $\Delta_{\rm {fs}}$. Relative $f^0$ intensities of
magnitude   $\sim 0.05$ for CeRhSi$_2$ and $\sim 0.07$ for Ce$_2$Rh$_3$Si$_5$ corresponds to $n_f$
value of the order 0.95 and 0.93, respectively. The Ce $3d$ XPS spectra, however, allow an estimate
of the occupation number $n_f$ (and of the energy $\Delta_{\rm {fs}}$) within an accuracy of the
order of 20\%. The errors due to the uncertainties in the intensity ratios we discussed previously
\cite{Slebarski2004}.

We note, however, that the $n_f$ value obtained from the Ce 3$d$ XPS spectra is smaller than that
estimated from the X-ray absorption spectra either for CeRhSi$_2$ or Ce$_2$Rh$_3$Si$_5$. This
discrepancy show immediately that the final-state $f$ occupation is a different function of the
initial-state occupancy for each type of spectroscopy (see Ref. \cite{Wuilloud83}). We note,
however, that the $n_f$ value for CeRhSi$_2$ is larger than $n_f$ derived for Ce$_2$Rh$_3$Si$_5$.

Other evidence for an fluctuating valence in Ce$_2$Rh$_3$Si$_5$ comes from the Ce 4$d$ XPS data. As
shown in Fig. 6, the latter spectrum exhibits some features above 120 eV (marked by the arrow),
which can be assigned to the $f^0$ final state \cite{Signorelli73,Baer78}.

\subsection{Electronic band structure}

The band structure calculations performed using two distinct approaches, i.e. with FPLO and
FP-LAPW, yielded very similar results and therefore only those obtained by using the Wien2k code
are discussed below. The latter method gives also an opportunity to go beyond LDA using L(S)DA+U
approach. The calculations were performed with and without spin polarization. The spin-polarized
calculations were started with finite initial magnetic moments in order to promote magnetic
solutions. However, the self-consistent results for magnetic solutions were nearly same as those
for nonmagnetic ones. In all the tests performed, the magnetic moments on cerium atoms were always
below 10$^{-7}$ $\mu_B$/Ce ~atom for both CeRhSi$_2$ and Ce$_2$Rh$_3$Si$_5$.

The calculated densities of states (DOS) for the compounds CeRhSi$_2$ and Ce$_2$Rh$_3$Si$_5$ are
presented in Figs. 7 and 8, respectively. In both cases the DOS plots can be divided into three
parts below the Fermi level (E$_F$ = 0). The first part located below -16~eV is composed of two
peaks formed by the Ce(5p$_{1/2}$) and Ce(5p$_{3/2}$) electrons. It should be noticed that the
LDA+U peaks are shifted towards higher binding energies. As will be discussed below, this shift
improves consistency between the calculated and measured X-ray photoelectron spectra, hence
justifying the LDA+U approach. The second part, located between about -6 and -11~eV, is formed
mainly by $s$ electrons of the Rh and Si atoms. The third part, which is the main one of the
valence bands, is located between the Fermi level and about -6~eV of binding energy. The main
contribution to these subbands is provided by $d$ electrons located on the Rh atoms and Si(3p)
electrons. A small contribution comes also from Ce(5d) electrons, but less than one electron per Ce
atom. The Ce(4f) electrons form a narrow band with the center of gravity above the Fermi level.
Most of the band is empty, and within the LDA scheme the number of occupied Ce(4f) states below
E$_f$ is equal to 0.92 for both systems. The LDA+U calculations change the electronic structure in
a way that the Ce(4f) band is slightly shifted above E$_F$ and consequently the number of Ce(4f)
electrons is reduced to 0.57 and 0.61 for CeRhSi$_2$ and Ce$_2$Rh$_3$Si$_5$, respectively. For the
former compound, the occupation of electrons inside the atomic sphere of Ce atom is given by the
LDA approach as 6s$^{1.98}$5d$^{0.64}$4f$^{0.92}$ and by the LDA+U calculations as
6s$^{1.99}$5d$^{0.81}$4f$^{0.57}$. These results should be compared with the configurations
6s$^{2}$5d$^{1}$4f$^{1}$ and 6s$^{2}$5d$^{2}$4f$^{0}$ of Ce$^{+3}$ and Ce$^{+4}$ ions,
respectively. Apparently, the calculated occupancies are intermediate between those characteristic
for stable valence ions. Similarly, the results were obtained for Ce$_2$Rh$_3$Si$_5$, which are
6s$^{1.98}$5d$^{0.67}$4f$^{0.92}$ and 6s$^{1.99}$5d$^{0.82}$4f$^{0.61}$ for LDA and LDA+U
calculations, respectively, also indicate distinct deviation from the stable valence configuration.
This effect may be caused by additional charge transfer from/to rhodium and silicon atoms as well
as by the charge accumulated in interstitial regions between atomic spheres.

The total DOS at E$_F$ and the main contributions of particular atoms are summarized in Table 3.
The main contributions, about 50\%, to the total DOS at E$_F$ are provided by Ce(4f) electrons.
Based on the DOS at E$_F$ one can calculate the Sommerfeld coefficient $\gamma_0$, to be equal to
about 11.6 and 11.2 mJ/(mol K$^2$) for CeRhSi$_2$ and Ce$_2$Rh$_3$Si$_5$, respectively, obtained
within the LDA approach and 3.7 and 6.5 mJ/(mol K$^2$), respectively, obtained with the LDA+U
approach. Comparison of these values to the experimental ones ($\gamma$ = 86 and 47 mJ/(mol K$^2$,
for the two compounds respectively) yields the mass enhancement factors
$\lambda$=($\gamma_{exp}$/$\gamma_0$)-1, which are also collected in Table 3.

Fig. 9 presents the X-ray photoelectron spectra of CeRhSi$_2$ and Ce$_2$Rh$_3$Si$_5$, calculated
from the DOS data using the cross sections reported in Ref. \cite{as11}. For comparison, the
experimental XPS spectra are also shown. Apparently, the calculated results are quite consistent
with the experimental ones. It should be noted that the LDA+U approach better describes the
measured spectra for both systems, as judged from the distinct differences in the region of the
Ce(5p) peaks.

\section{Summary}

The macroscopic and spectroscopic results obtained in this work for CeRhSi$_2$ and
Ce$_2$Rh$_3$Si$_5$ unambiguously corroborate intermediate valent character of both compounds.
Mutually comparing the properties of these two phases it is worth noting that the effective valence
of Ce ion in CeRhSi$_2$ is closer to +3 than in the other silicide in the entire temperature range
studied. In both compounds the electronic ground state of cerium is $4f^0$, yet the energy distance
to the excited magnetic $4f^1$ state, $E_{\rm {ex}}$, is smaller in CeRhSi$_2$ and hence the
magnetic susceptibility of this compound is larger and more temperature dependent than the
susceptibility of the other one. Furthermore, stronger hybridization of the $4f$ electronic states
with the conduction states in Ce$_2$Rh$_3$Si$_5$ is clearly reflected in larger values of the
characteristic spin and charge fluctuation temperatures $T_0$ and $T_{\rm {sf}}$. Similar
conclusions as regards the valence states in the two compounds can be derived from the XAS and XPS
results. The mean occupation of the 4$f$ state is smaller in Ce$_2$Rh$_3$Si$_5$ and less
temperature dependent than in the other silicide, in line with the bulk magnetic data. The
experimental XPS spectra can be fairly well reproduced by ab-initio electronic band structure
calculations, which yielded nonmagnetic ground state in both compounds with significant mass
enhancements due to strong electronic correlations.

\ack {The authors are grateful to Dr. Marek Wo{\l}cyrz for collecting the X--ray powder diffraction
data. This work was supported by the Ministry of Science and Higher Education within the research
projects No. N202 116 32/3270 and N202 1349 33. Part of the research was performed in the frame of
the National Network "Strongly correlated materials: preparation, fundamental research and
applications"}

\newpage
\begin{table}
\caption{\label{tab1} Atomic positions and equivalent isotropic thermal displacement parameters for
CeRhSi$_2$.}
\begin{center}
\begin{tabular}{lccccc}
Atom & Site & x & y & z & U(eq)\\
\hline
Ce & 4c & 0 & 0.8945(1) & 0.75 & 0.0108(6)\\
Rh & 4c & 0.5 & 0.8211(1) & 0.25 & 0.037(1)\\
Si1 & 4c & 0 & 0.5427(7) & 0.75 & 0.032(4)\\
Si2 & 4c & 0.5 & 0.7477(5) & 0.75 & 0.010(3)\\
\end{tabular}
\end{center}
\end{table}

\begin{table}
\caption{\label{tab2} Atomic positions and equivalent isotropic thermal displacement parameters for
Ce$_2$Rh$_3$Si$_5$.}
\begin{center}
\begin{tabular}{lccccc}
Atom & Site & x & y & z & U(eq)\\
\hline
Ce & 8j & 0.2668(1) & 0.1305(1) & 0 & 0.0072(3)\\
Rh1 & 4a & 0 & 0 & 0.75 & 0.0097(6)\\
Rh2 & 8j & 0.3923(1) & 0.1386(2) & 0.5 & 0.0108(4)\\
Si1 & 4b & 0.5 & 0 & 0.75 & 0.018(2)\\
Si2 & 8g & 0 & 0.2218(4) & 0.75 & 0.017(2)\\
Si3 & 8j & 0.1574(6) & 0.1095(6) & 0.5 & 0.021(2)\\
\end{tabular}
\end{center}
\end{table}

\begin{table}
\caption{\label{tab3} DOS at E$_F$: total (per eV and f.u.) and site projected (per eV and atom)
from orbitals giving the main contributions to the total value; calculated Sommerfeld coefficients
$\gamma_0$ in mJ/(mol K$^2$) and mass enhancement factors $\lambda$.}
\begin{center}
\begin{tabular}{lccccccc}
Compound & Type od & Total & $\gamma_0$ & $\lambda$ & Ce(4f) & Rh1(4d)
& Rh2(4d)\\
& calculations & & & & & & \\
\hline
CeRhSi$_2$ & LDA & 4.92 & 11.60 & 6.4 & 2.87 & 0.32 & - \\
& LDA+U & 1.58 & 3.72 & 22.1 & 0.67 & 0.16 & - \\
Ce$_2$Rh$_3$Si$_5$ & LDA & 4.73 & 11.15 & 3.2 & 1.19 & 0.55 & 0.19 \\
& LDA+U & 2.74 & 6.46 & 6.3 & 0.61 & 0.27 & 0.15 \\
\end{tabular}
\end{center}
\end{table}

\newpage

\begin{figure}[t]
\centering
{\includegraphics{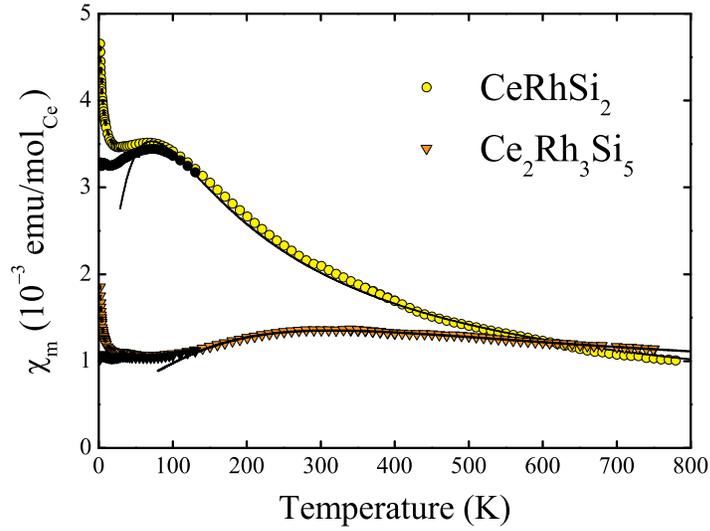}}
\caption{Temperature dependences of the molar magnetic
susceptibility of CeRhSi$_2$ and Ce$_2$Rh$_3$Si$_5$. The full
symbols represent the $\chi(T)$ curves corrected for contributions
coming from spurious Ce$^{3+}$ ions (see text for details). The
dashed and solid lines are the fits discussed in the text.}
\end{figure}

\begin{figure}[t]
\centering
{\includegraphics{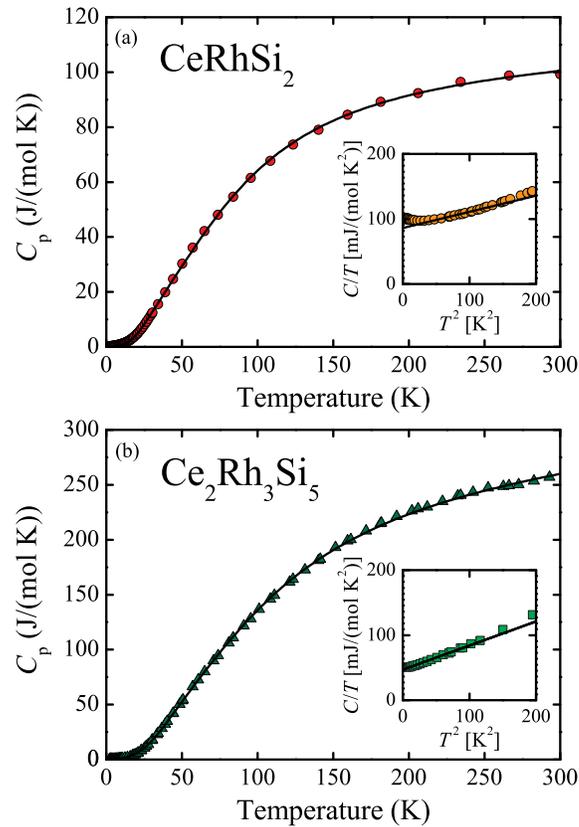}}
\caption{Temperature dependences of the specific heat of (a)
CeRhSi$_2$ and (b) Ce$_2$Rh$_3$Si$_5$. The insets present the
low-temperature data in the form $C/T$ vs. $T^2$. The solid lines
are the fits discussed in the text.}
\end{figure}

\begin{figure}[t]
\centering
{\includegraphics{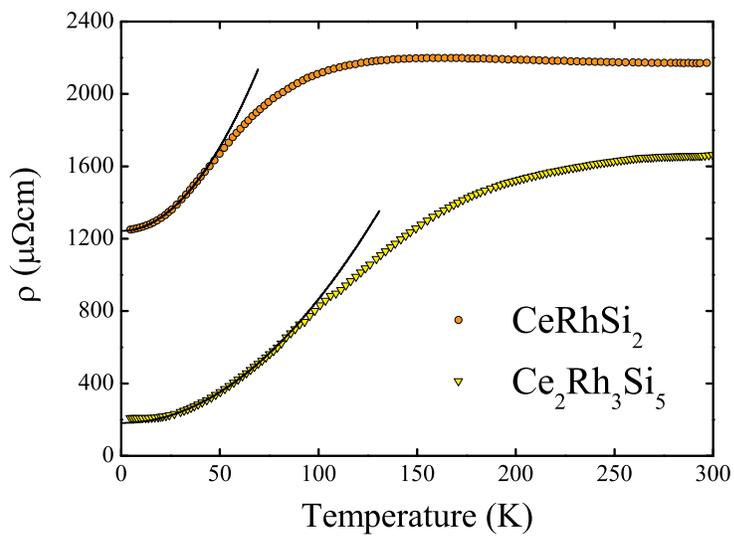}}
\caption{Temperature dependences of the electrical resistivity of
CeRhSi$_2$ and Ce$_2$Rh$_3$Si$_5$. The solid lines are the fits
discussed in the text.}
\end{figure}

\begin{figure}[t]
{\includegraphics{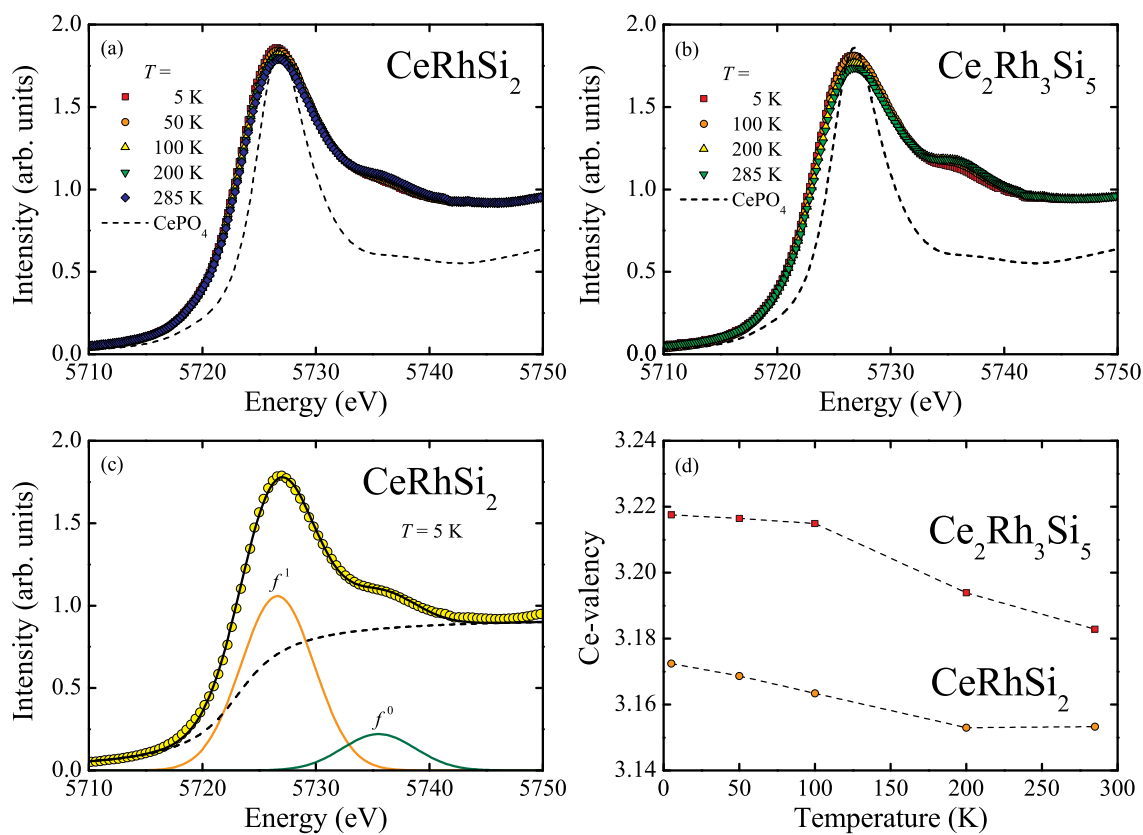}}
\centering
\caption{X-ray absorption spectra at the Ce-L$_{\rm III}$ threshold measured at various
temperatures for (a) CeRhSi$_2$ and (b) Ce$_2$Rh$_3$Si$_5$. The standard spectrum for Ce$^{3+}$
ions, i.e. that of CePO$_4$ is represented by the dashed curves. As an example, deconvolution of the
spectrum measured for CeRhSi$_2$ at 5 K is shown in panel (c). The contributions due to 4$f^1$ and
4$f^0$ configurations are depicted by the solid lines, while the dashed line represents the
arctan-function accounting for transitions of the $2p$ electrons to the
conduction band. Panel (d) displays the calculated changes of the Ce ions valence in both compounds
with varying the temperature.}
\end{figure}

\begin{figure}[t]
\centering
{\includegraphics{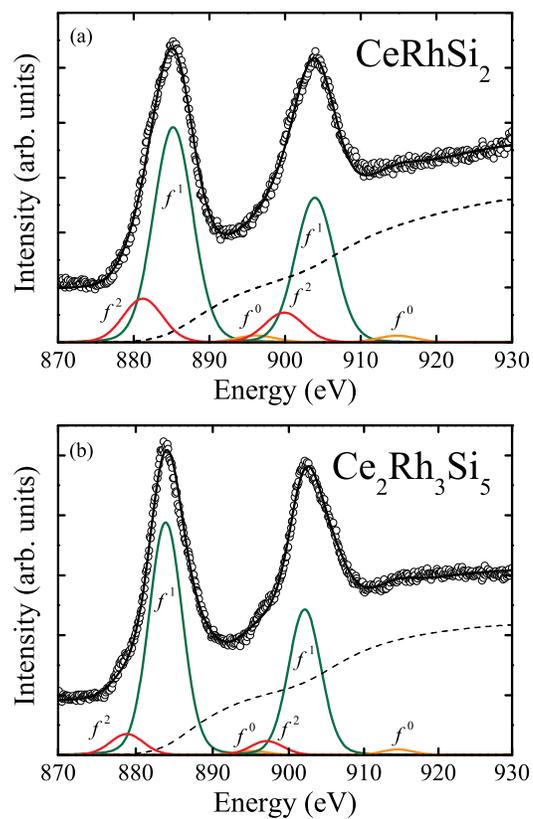}}
\caption{Ce 3\textit{d} XPS spectra obtained for (a) CeRhSi$_2$ and (b) Ce$_2$Rh$_3$Si$_5$. The
$f^{0}$, $f^{1}$ and $f^{2}$ components were separated on the basis of the Doniach-\v
Sunji\'c theory.}
\end{figure}

\begin{figure}[t]
\centering
{\includegraphics{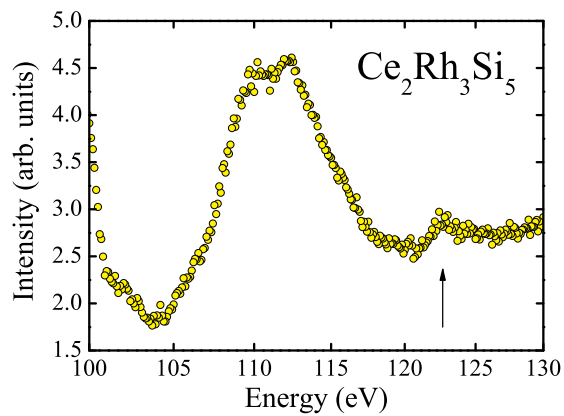}}
\caption{Ce 4\textit{d} XPS spectrum obtained for Ce$_2$Rh$_3$Si$_5$.}
\end{figure}

\begin{figure}[t]
\centering
{\includegraphics{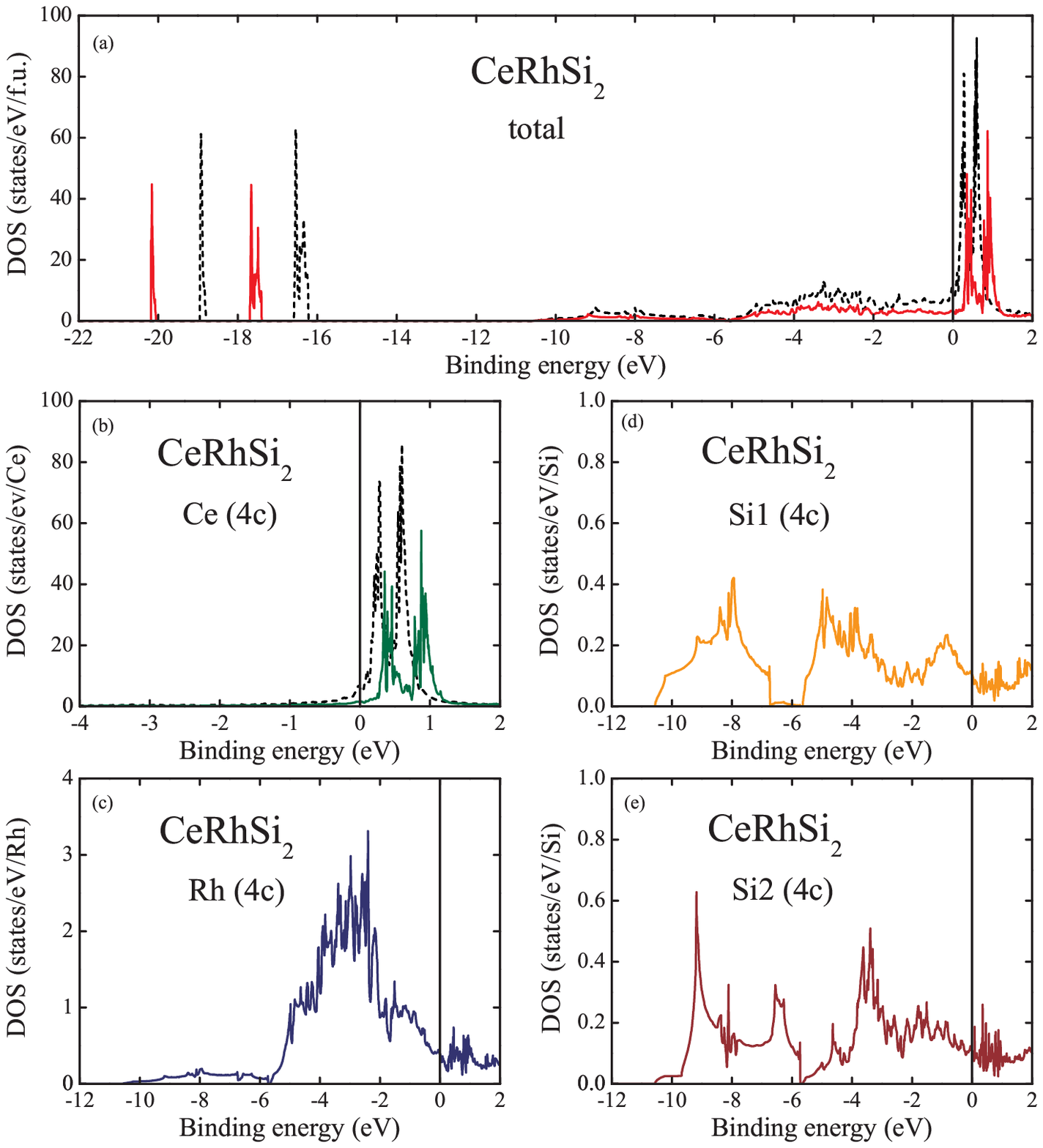}}
\caption{The total (per f.u.) and site projected (per atom)
densities of electronic states (DOS) for CeRhSi$_2$ calculated
within the LDA+U scheme (for comparison the total DOS and the Ce-atom DOS
calculated using LDA approach are presented by the broken line).}
\end{figure}

\begin{figure}[t]
\centering
{\includegraphics{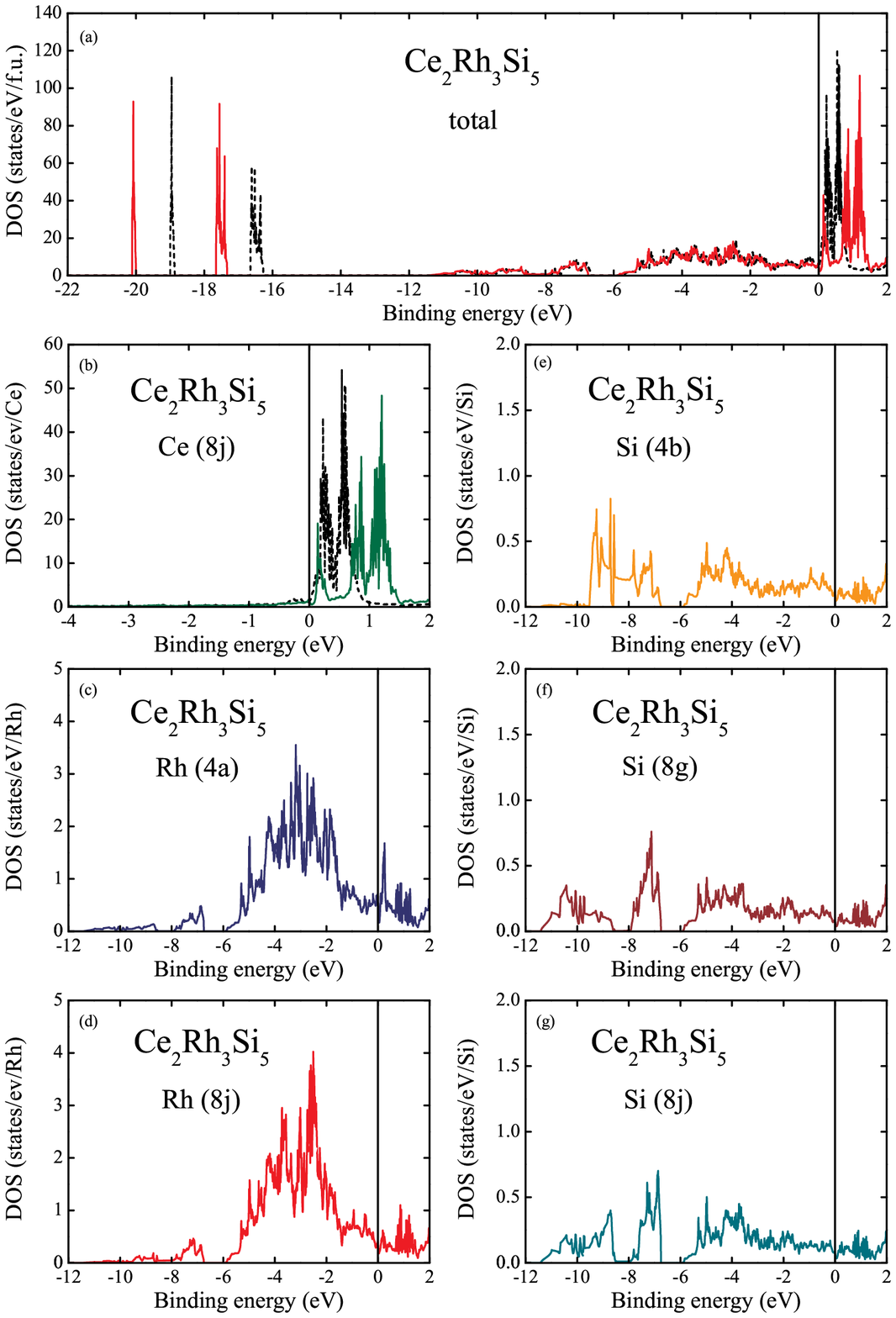}}
\caption{The total (per f.u.) and site projected (per atom)
densities of electronic states (DOS) for Ce$_2$R$_3$hSi$_5$
calculated within the LDA+U scheme (for comparison the total DOS and the Ce-atom DOS
calculated using LDA approach are presented by the broken line).}
\end{figure}

\begin{figure}[t]
\centering
{\includegraphics{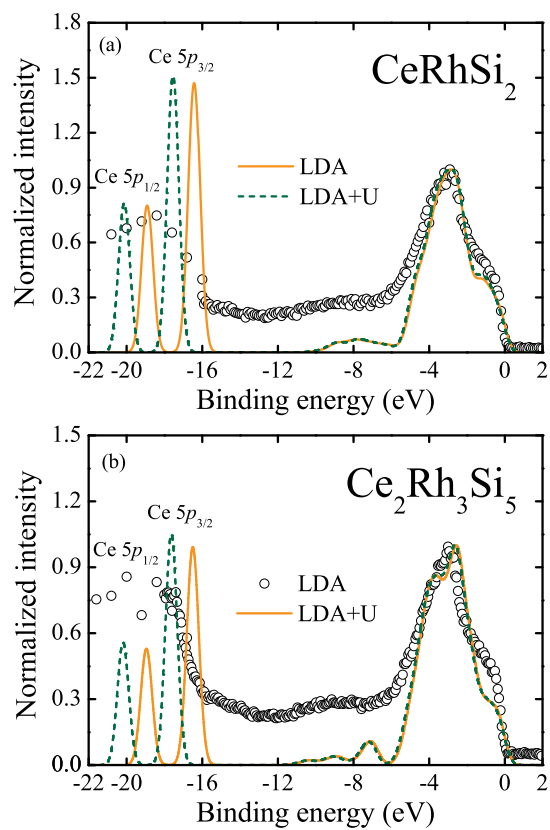}}
\caption{XPS valence band spectrum of (a) CeRhSi$_2$ and (b) Ce$_2$Rh$_3$Si$_5$, calculated
within the LDA and LDA+U approaches and compared to the experimental one (represented by the
symbols). Note the difference between the calculated spectra in the region of 5$p$
contribution.}
\end{figure}

\end{document}